\documentclass[aps,prb,superscriptaddress,showpacs,twocolumn]{revtex4}
\usepackage{times,mathptm}
\begin{document}
\title{Substrate effects in magneto-optical second-harmonic generation
  from first principles: Fe/Cu(001)}
\author{Torsten Andersen}
\email{thor@mpi-halle.mpg.de}
\affiliation{Max-Planck-Institut f{\"u}r Mikrostrukturphysik, Weinberg
  2, D--06120 Halle/Saale, Germany}
\author{W.~H{\"u}bner}
\email{huebner@mpi-halle.mpg.de}
\affiliation{Max-Planck-Institut f{\"u}r Mikrostrukturphysik, Weinberg
  2, D--06120 Halle/Saale, Germany}
\affiliation{Institut f{\"u}r Theoretische Physik,
  Karl-Franzens-Universit{\"a}t Graz, Universit{\"a}tsplatz 5, 
  A--8010 Graz, Austria}

\begin{abstract}
  We compute the nonlinear optical response of an Fe monolayer placed
  on top of 1 to 4 monolayers of Cu(001). Our calculation is based on
  {\it ab initio}\/ eigenstates of the slab, which are obtained within
  the full-potential linearized augmented plane-wave method. The
  ground-state spin-polarized electronic structure is converged
  self-consistently to an accuracy better than 0.1 mRy. Subsequently,
  we take the spin-orbit interaction into account within a second
  variational treatment. The new set of eigenstates allows us to
  calculate the magneto-optical transition matrix elements. The
  second-harmonic response is determined in the reflection geometry
  with magnetization perpendicular to the surface (the so-called polar
  configuration) using the surface-sheet model. Adding layers of a
  noble metal (Cu) to the Fe monolayer gives a new degree of freedom
  for the inclusion of nonmagnetic Cu d bands to the nonlinear
  magneto-optical response of the slab, and the energy bands show that
  such an addition converges essentially to an addition of d-states
  and a small broadening of the d-band with growing number of Cu
  layers. The screened nonlinear optical susceptibility is calculated
  and converges quite well with a growing number of Cu layers. Our
  first-principles results confirm that the magnetic tensor elements
  of the nonlinear optical response tensor are roughly of the same
  order of magnitude as the nonmagnetic ones (in contrast to linear
  optics, where the magnetic response is only a minor correction).
\end{abstract}
\maketitle

\end{document}